\documentstyle[12pt, psfig]{article}
\renewcommand{\thepage}{\arabic{page}}
\newcommand{\nc}{\newcommand}
\nc{\beq}{\begin{equation}} \nc{\eeq}{\end{equation}}
\nc{\beqa}{\begin{eqnarray}} \nc{\eeqa}{\end{eqnarray}}
\nc{\lsim}{\begin{array}{c}\,\sim\vspace{-21pt}\\< \end{array}}
\nc{\gsim}{\begin{array}{c}\sim\vspace{-21pt}\\> \end{array}}

\newcommand{\drawsquare}[2]{\hbox{%
\rule{#2pt}{#1pt}\hskip-#2pt%  left vertical
\rule{#1pt}{#2pt}\hskip-#1pt%  lower horizontal
\rule[#1pt]{#1pt}{#2pt}}\rule[#1pt]{#2pt}{#2pt}\hskip-#2pt%  upper horizontal
\rule{#2pt}{#1pt}}% right vertical

\newcommand{\Yfund}{\raisebox{-.5pt}{\drawsquare{6.5}{0.4}}}%  fund
\newcommand{\Ysymm}{\raisebox{-.5pt}{\drawsquare{6.5}{0.4}}\hskip-0.4pt%
        \raisebox{-.5pt}{\drawsquare{6.5}{0.4}}}%  symmetric second rank
\newcommand{\Yasymm}{\raisebox{-3.5pt}{\drawsquare{6.5}{0.4}}\hskip-6.9pt%
        \raisebox{3pt}{\drawsquare{6.5}{0.4}}}%  antisymmetric second rank

\newcounter{mysection}
\newcounter{mysubsection}
\newcommand{\mysection}[1]{\stepcounter{mysection}\setcounter{equation}{0}
\setcounter{mysubsection}{0}\par\bigskip\noindent{\large\bf
\themysection .\ #1}\nopagebreak[4]\par\vskip .3cm}

\newcommand{\mysubsection}[1]{\stepcounter{mysubsection}
\par\medskip\noindent{\large\it
\themysection .\themysubsection\ #1}\nopagebreak[4]\par\vskip .3cm}
\newcommand{\mysectionstar}[1]{
\par\bigskip\noindent{\large\bf #1}\nopagebreak[4]\par\vskip .3cm}
\def\l:{\mathopen{:}\,}
\def\r:{\,\mathclose{:}}

\def\inbar{\,\vrule height1.5ex width.4pt depth0pt}
\font\cmss=cmss12 \font\cmsss=cmss12 at 7pt
\def\IZ{\relax\ifmmode\mathchoice
{\hbox{\cmss Z\kern-.4em Z}}{\hbox{\cmss Z\kern-.4em Z}}
{\lower.9pt\hbox{\cmsss Z\kern-.4em Z}}
{\lower1.2pt\hbox{\cmsss Z\kern-.4em Z}}\else{\cmss Z\kern-.4em
Z}\fi}
\def\IB{\relax{\rm I\kern-.18em B}}
\def\IC{{\relax\hbox{$\inbar\kern-.3em{\rm C}$}}}
\def\ID{\relax{\rm I\kern-.18em D}}
\def\IE{\relax{\rm I\kern-.18em E}}
\def\IF{\relax{\rm I\kern-.18em F}}
\def\IG{\relax\hbox{$\inbar\kern-.3em{\rm G}$}}
\def\IP{\relax{\rm I\kern-.18em P}}
\begin{document}

\begin{titlepage}

{\hbox to\hsize{hep-ph/9712193 \hfill Fermilab-Pub-97/425-T}}
{\hbox to\hsize{December 1997 \hfill UCSD-PTH-97-37}}
\bigskip

\begin{center}

\vspace{.5cm}

\bigskip

\bigskip

\bigskip

{\Large \bf   M(ore) on  Chiral Gauge Theories from D-Branes}

\bigskip

\bigskip

{\bf Joseph Lykken}$^{\bf a}$, 
{\bf Erich Poppitz}$^{\bf b}$, and {\bf Sandip P. Trivedi}$^{\bf a}$ \\

\smallskip

{\tt lykken@fnal.gov, epoppitz@ucsd.edu, trivedi@fnal.gov}

\bigskip

\bigskip

$^{\bf a}${ \small \it Fermi National Accelerator Laboratory\\
  P.O.Box 500\\
 Batavia, IL 60510, USA\\}

\bigskip

$^{\bf b}${\small \it Department of Physics \\
University of California at San Diego\\
9500 Gilman Drive\\
La Jolla, CA 92093, USA}
 
\bigskip
 
\bigskip

{\bf Abstract}

\end{center}
We consider a brane configuration consisting of intersecting Neveu-Schwarz
five-branes, Dirichlet four-branes, and an orientifold four-plane in a
$\IC^2/\IZ_3$ orbifold background. We show that the low-energy dynamics
is described by a four dimensional gauge theory with $N=1$ supersymmetry and 
$SO(N+4) \times SU(N)$ or $SP(2M) \times SU(2M+4)$ gauge symmetry.
The matter content of this theory is chiral. In particular,   the $SU$ group has  one 
matter  field in the  antisymmetric tensor or symmetric  tensor representation and 
several fields in the fundamental  and antifundamental representations.
We discuss various consistency checks  on these theories. By 
considering the brane configuration in M theory  we deduce the spectral curves for
these theories. Finally, we consider the effects of replacing the orbifold background
with a non-singular ALE space (both with and without an orientifold plane) and
show that it leaves the spectral curves unchanged.

\end{titlepage}

\renewcommand{\thepage}{\arabic{page}}

\setcounter{page}{1}

\mysection{Introduction.}

Lately, many insights into gauge theories have emerged from the study
of D-branes \cite{vafa1}--\cite{LW}.  
Two approaches have proved useful for this purpose.  
One, involving studying branes in Calabi-Yau backgrounds, was
pioneered in \cite{vafa1}, for a review see \cite{geomeng} 
and references therein. 
Another is to consider configurations of intersecting D-branes and NS
branes \cite{hw}. As noted in \cite{wittenone}, such configurations often
correspond in M theory to a single smooth NS brane and 
this observation allows one to deduce the quantum behavior  
of the resulting gauge theory 
from classical considerations in M theory \cite{hw}--\cite{LW}.  

In a previous paper \cite{lpt}, we showed how this latter approach could be
extended  to  study  some chiral gauge theories.  The key idea was to consider 
intersecting brane configurations in non-trivial backgrounds. 
Ref.~\cite{lpt}  
focused,
in particular, on   orbifold  backgrounds. In this  
paper  we    continue the study  by considering   
brane-configurations in a $\IZ_3$
orientifold background \footnote{Generalising these results  to $\IZ_N$
backgrounds is straightforward but the  resulting gauge theories are more
complicated and not particularly illuminating. We  will  not  study  them  
 here.}. This yields a rich class of  four dimensional 
 theories with  $N=1$ supersymmetry,    $SU (N)\times SO(N+4)$ or  
$SU(2M+4) \times SP(2M)$ gauge symmetry and matter in chiral
representations.  In particular the $SU$ group  contains one field in the
antisymmetric  or symmetric
representation and several fields in the  fundamental and antifundamental
representations.  The analysis leading to the field theory is similar to the orbifold
case in many respects but has one new feature: 
 the orientifold  carries a  charge with respect to  twisted RR fields which
must be cancelled by D
4-branes.  We show how the requirement of anomaly cancellation fixes
this charge  uniquely   and determines  the field theories. 

In Section 3, we study the classical moduli space of the brane configuration 
and show that it agrees with the field theory analysis.
In  Section 4, we  turn to considering the brane-configuration in  M theory
and show how various  non-perturbative features of the  low-energy
dynamics, pertaining to the spectral curves , can be  derived in this way. 
Finally, in Section 5,  we return to the orbifold theories considered in 
\cite{lpt}
and  show that  the spectral curves  are left unchanged (in suitable
coordinates) when the orbifold  background is replaced by a non-singular 
ALE space.  This is consistent with a field theory argument showing that the 
curves should be independent of the Fayet-Iliopoulos terms for the anomalous
$U(1)$'s. We
also make some comments with regards to  orientifolds in this  context.

One main motivation behind this work has been to gain a better 
understanding of chiral gauge theories which exhibit
supersymmetry breaking. 
We hope some of the techniques
developed here will eventually prove useful for this purpose.

\mysection{$\IZ_3$ Orientifolds.}

\mysubsection{The orientifold  and brane configuration.}

In this paper   we will  discuss   a  particular
 orientifold background  with an   orientifold group  given by
\beq
\label{orgroup}
G~=~\{~1, ~\alpha, ~\alpha^2 ,~ \Omega  R  (-1)^{F_L} ,
~ \Omega R (-1)^{F_L} \alpha,~
 \Omega R (-1)^{F_L} \alpha^2 ~\}~. 
\eeq 
Here, 
$\alpha$  is a spacetime symmetry that  acts on  $v=X^4+i X^5$ and $w=X^8+i
X^9$ as  
\beq
\label{defzthree}
(v, w)  \rightarrow ( \alpha   v, \  \alpha^{-1}  w), ~\alpha 
~\equiv ~e^{{2 \pi i  \over 3}},
\eeq
$R$  is   a reflection which acts on the directions transverse to the 
orientifold 4-plane, $X^4, X^5,$ $ X^7, X^8, X^9$, and $\Omega$  denotes 
world-sheet orientation reversal. The $\IZ_6$ orientifold group has 
a $\IZ_2$ subgroup:
\beq
\label{defz2}
g_1~=~ \{~1,~ \Omega R (-1)^{F_L} \},
\eeq
and a $\IZ_3$ subgroup
\beq
\label{defz3}
g_2~=~ \{~1, ~\alpha,~ \alpha^2 \}.
\eeq
In terms of these, the projection  corresponding to the orientifold
is given by 
\beq
\label{defproj}
\left( {1 + \Omega  R  (-1)^{F_L} \over 2} \right) 
~\left( {1 + \alpha + \alpha^2 \over 3} \right)~.  
\eeq
It is useful to bear in mind that the orientifold
described above is related by  
T-duality to a $\IC^2/{\IZ_3}$ orbifold of  Type I theory\footnote{In 
this paper we  consider  both $SO$ and $SP$ orientifolds; the  above 
remark applies to the $SO$ case. Also  we  only
consider non-compact backgrounds  here; strictly speaking the remark on
T-duality  applies to a background that has been
appropriately compactified.}.  

We will consider a configuration of intersecting NS and D4-branes placed in
this  background\footnote{While, as 
mentioned above, the generalization to arbitrary abelian ($\IC^2/\IZ_n$) 
orbifolds is straightforward, the nonabelian case
(i.e. $\IC^2/G$, with $G$--a nonabelian discrete subgroup of $SU(2)$)
of the type considered in \cite{DM}, \cite{BI2}, is 
problematic. This is because in
the construction with NS branes of Fig.1 a nonabelian orbifold
would mix parallel and transverse
directions to the NS branes.}
as shown in Fig. 1. The NS branes, stretching
along $X^1, X^2, X^3, X^4, X^5$, 
are placed at the orientifold point, $X^7=X^8=X^9=0$, and 
have definite positions in $X^6$. The D4-branes
stretch along $X^1, X^2, X^3, X^6$  and are also placed at
$X^7=X^8=X^9=0$, which ensures that they intersect the NS branes.  Finally,
there
is an orientifold  4-plane (O4-plane) parallel to the D4-branes (along 
$X^1, X^2, X^3, X^6$), which is also placed at  the
orientifold fixed point, $X^4=X^5=X^7=X^8=X^9=0$,
and also intersects the NS branes. 

\begin{figure}[ht]
\vspace*{13pt}
\centerline{\psfig{file=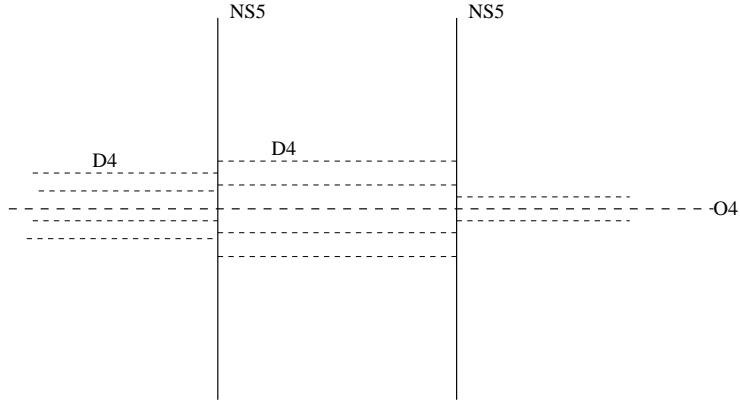}}
\vspace*{13pt}
\caption{The brane configuration giving rise to $N=2$ $SO$ or 
$SP$ theories.}
\end{figure}

\mysubsection {The gauge group and matter content: basic strategy.}

We  are  interested in studying the low-energy dynamics of this brane
configuration.  By an argument analogous to the $N=2$ case,
it follows that 
this dynamics  is  governed by a $3+1$ dimensional  theory living 
in the intersection region of the NS and the D 4-branes, i.e., 
along $X^1, X^2, X^3$.   The  brane configuration  in Fig.~1 leaves  
four supercharges unbroken, corresponding to $N=1$ supersymmetry in $3+1$
dimensions.  This  follows by noting that  in the absence of the
orientifold   the configuration   is invariant under  eight supercharges;
``turning on" the  orientifold  reduces this number   to four.  

The main question that will occupy us in this section has to do with 
the gauge  group and matter  content of this field theory. Since the
discussion  is somewhat involved  we  will first sketch out the basic 
strategy  in this section. Following this,  in section 2.3, the strategy will 
be  explicitly implemented  to find 
the resulting gauge group and matter content. 

It is  useful at the outset to summarize some conclusions
from our earlier study 
of the field theories obtained by placing the brane configuration in Fig.~1 in
orbifold backgrounds, \cite{lpt}.
In these cases, it was found that the resulting
field theory  could be obtained  by  starting with    
an $N =2$  supersymmetric theory and projecting out some states.  
The projection involved two operations. One 
was a spacetime symmetry (dictated
by the particular orbifold background)   
and the second acted on the Chan-Paton
indices. States  which
transformed  non-trivially under this combined  
transformation were projected out, while
those that were  invariant survived and gave rise to the low-energy theory.
This projection could also  be understood in  purely field 
theoretic terms  in the underlying $N=2$
theory. To see this, recall that  the space-time symmetries of the $N=2$
brane configuration, $SO(3)_{7,8,9} \times SO(2)_{4,5}$, appear as 
the classical 
$SU(2)\times U(1)$ R symmetry of the $N=2$ world volume theory.
The orbifold spacetime symmetry then   
corresponded to  a  non-anomalous  discrete 
subgroup of the R symmetries\footnote{To avoid confusion, 
we stress that the relevant
combination of R symmetries is non-R in $N=1$ terms, i.e. its action on all
states in $N=1$ supermultiplets is identical.}, while 
the  action on the Chan-Paton factors corresponded
to  a discrete subgroup of the gauge symmetry. 

For the  orientifold background  at hand, we  also 
expect that the gauge group and matter content can be obtained by  starting 
with an  $N=2$ supersymmetric  theory and   truncating  the spectrum. 
The $\IZ_2$ subgroup  of the orientifold group, eq.~(\ref{defz2}),
suggests that   the $N=2$ theory has an $SO$ or $SP$ gauge symmetry.   
As in the orbifold case above, the  projection operator  used  in truncating 
the spectrum should have  an interpretation  in the field theory  
as the  product of  a discrete  subgroup of  the R symmetries  
and  a discrete  subgroup  of  the  gauge symmetry. 

The discrete subgroup of the R symmetry 
corresponds to the spacetime symmetry,
eq.~(\ref{defzthree}), and can be easily found. 
  Identifying the appropriate subgroup of the gauge symmetry is  
  trickier  and will take more effort.  This subgroup corresponds to the action
on  the Chan-Paton  indices. 
The action of the group elements,
$\Omega R (-1)^{F_L}$ and $\alpha$, eq.~(\ref{orgroup}), on the  Chan-Paton
factors $\lambda$ can be  represented by the matrices :  
\beq
\label{eomega}
\lambda \rightarrow \gamma_{\Omega R (-1)^{F_L}} ~ \lambda^T ~
\gamma_{\Omega R (-1)^{F_L}}^{-1}~,
\eeq
and 
\beq
\label{egamma} \lambda \rightarrow \gamma_\alpha ~ \lambda ~
\gamma_\alpha^{-1}.  
\eeq 
The matrices   $\gamma_\alpha $ and  $\gamma_{\Omega R (-1)^{F_L}}$ must  
furnish a
representation of the orientifold group. From  eq.~(\ref{orgroup}) 
this implies, 
\beq
\label{orgroupb}
\gamma_{\Omega R (-1)^{F_L}}\ = \ 
\pm \ \left( \gamma_{\Omega R (-1)^{F_L}} \right)^{T}
\eeq 
\beq
\label{orgroupa}
\gamma_\alpha^3= {\rm phase} .
\eeq
The  $+$ $(-)$  choice in  eq.~(\ref{orgroupb})
 corresponds to starting   with an $N=2$  theory   with $SO$($SP$)  symmetry. 
Eq.~(\ref{orgroupa}) restricts the   subgroup of the gauge symmetry 
involved in the truncation  but does not specify it uniquely. 
The remaining ambiguity in this choice is usually fixed by the requirement of 
tadpole cancellation \cite{GP} for 
twisted RR fields\footnote{For the orientifold 
discussed here, the twisted RR fields
involved can propagate along one spatial
direction, $X^7$, which is  of infinite extent. 
Even so, as argued  in ref.~\cite{BI}, one
expects  the tadpoles to cancel.  
For our purposes this argument is suggestive; however,  we
actually  use anomaly cancellation in the field theory  to provide us  with an
additional constraint.}.  
The tadpole cancellation condition ensures  that the total charge coupling to
the twisted RR field vanishes.  
In the present context, the  RR fields come from the twisted
sectors corresponding to the $(1,\alpha,\alpha^2)$ subgroup of the orientifold
group.  Since  both the D4-branes and the orientifold plane 
can  carry  charge  for these RR fields, the tadpole cancellation 
condition should
provide  a constraint of the form:  
\beq
\label{orcharge}
{\rm Tr} \gamma_\alpha \ = \ - \ {\rm orientifold \  charge} \ .
\eeq
The RR tadpoles are usually
determined by one loop calculations, but with NS branes present no explicit
conformal field theory is known  and such a calculation poses a formidable
challenge.   

How then are we to proceed in identifying the 
discrete subgroup of the  gauge
symmetry involved in projecting out states?  
As was discussed in ref.~\cite{CP}  and more
recently refs.~\cite{GJ}, \cite{DP}, one physical consequence of imposing  
tadpole cancellation  is 
to ensure that the low-energy field theory is anomaly free.   
Accordingly, we  will, to begin with, 
leave the orientifold charge, eq.~(\ref{orcharge}), to be a free parameter.
Starting with an $SO$ or $SP$  $N=2$ theory we  truncate the spectrum,
using 
the discrete subgroup of the R symmetry and $\gamma_\alpha$,
and then
demand that the resulting field  theory is anomaly free. 
As we shall see, this fixes the orientifold charge  
uniquely  and  completely determines the field theory.  

It is worth pointing out that, in  
general,  anomalies on a brane  world volume theory  do not 
have to cancel.   When  this happens,
the corresponding
non-conservation of  charge  is due to a charge inflow 
from outside the brane \cite{CH}, \cite{BH}. 
In our case, however,
the  gauge fields involved vanish outside the brane  world
volume. The  non-abelian  gauge anomalies in the world volume theory
must therefore vanish. 

As we see below, the $N=2$, $SO$ theory  leads to a theory with $SO(N+4) 
\times U(N)$ gauge symmetry and $SP(2M_L) \times U(2M_L+4) \times 
SP(2M_R) \times U(2M_R+4)$  global  symmetry. The matter content
of this theory is given by: 
\begin{equation}
\label{sogauged}
 \begin{array}{cc|c|c|c}
 &    &   global    &   gauge   &   global  \\
   &  & SP(2 M_L) \times U(2 M_L + 4) & SO(N +4) \times U(N) & 
      SP(2 M_R) \times U(2 M_R + 4) \\ \hline
4-4& \bar{Q} & 1 \hspace{.8 cm} 1  & \Yfund  \hspace{.8 cm} \overline{\Yfund}  
& 1  \hspace{.8 cm} 1 \\
strings& A & 1  \hspace{.8 cm} 1       & 1 \hspace{.8 cm} \Yasymm 
 &1 \hspace{.8 cm} 1    \\ \hline
left& v_L &1 \hspace{.8 cm} \overline{\Yfund}  &  \Yfund \hspace{.8 cm} 1 
& 1 \hspace{.8 cm} 1 \\
strings& q_L & 1 \hspace{.8 cm}\Yfund &  1 \hspace{.8 cm} \Yfund  
& 1 \hspace{.8 cm} 1 \\
& \bar{q}_L &  \Yfund \hspace{.8 cm} 1& 1 \hspace{.8 cm}
\overline{\Yfund} 
 & 1 \hspace{.8 cm} 1 \\ \hline
right & v_R &  1 \hspace{.8 cm} 1  &   \Yfund \hspace{.8 cm} 1  
& 1 \hspace{.8 cm} \overline{\Yfund} \\
strings& q_R & 1 \hspace{.8 cm} 1   & 1 \hspace{.8 cm} \Yfund  
 & 1 \hspace{.8 cm} \Yfund\\
& \bar{q}_R &  1 \hspace{.8 cm} 1   &1 \hspace{.8 cm}
\overline{\Yfund}  
& \Yfund \hspace{.8 cm} 1
    \end{array} \nonumber
\end{equation}

\smallskip

The $N=2$, $SP$ theory, on the other hand,
will give rise to  theories with  $SP(2M) \times U(2M+4)$
gauge symmetry  and $SO(N_L+4) \times U(N_L) \times SO(N_R+4) \times 
U(N_R)$  global symmetry. The matter content is given by: 
  \begin{equation}
\label{spgauged}
 \begin{array}{cc|c|c|c}
   &   &  global  &  gauge  &  global  \\
   &  & SO(N_L + 4) \times U(N_L) & SP(2 M) \times U(2 M + 4) & 
      SO(N_R + 4 ) \times U(N_R ) \\ \hline
  4-4  & Q  & 1 \hspace{.8 cm} 1  & \Yfund  \hspace{.8 cm}\overline{\Yfund}   
  & 1  \hspace{.8 cm} 1 \\
  strings  & S & 1  \hspace{.8 cm} 1  & 1 \hspace{.7 cm}
  \Ysymm 
   &1 \hspace{.8 cm} 1    \\ \hline
  left & v_L & \Yfund \hspace{.8 cm} 1 & 1 \hspace{.8 cm}\overline{\Yfund} 
  & 1 \hspace{.8 cm} 1 \\
  strings& q_L & 1 \hspace{.8 cm} \Yfund & 1 \hspace{.8 cm}\Yfund 
  & 1 \hspace{.8 cm} 1 \\
  & \bar{q}_L & 1 \hspace{.8 cm} \overline{\Yfund} & \Yfund
\hspace{.8 cm} 1
   & 1 \hspace{.8 cm} 1 \\ \hline
 right & v_R &  1 \hspace{.8 cm} 1  & 1 \hspace{.8 cm}\overline{\Yfund} 
 &   \Yfund \hspace{.8 cm} 1 \\
  strings& q_R & 1 \hspace{.8 cm} 1 & 1 \hspace{.8 cm}\Yfund 
  & 1 \hspace{.8 cm} \Yfund \\
  & \bar{q}_R &  1 \hspace{.8 cm} 1  & \Yfund \hspace{.8 cm} 1 
  &1 \hspace{.8 cm} \overline{\Yfund} 
    \end{array} \nonumber
\end{equation}

\smallskip

Note that  the matter content  in (\ref{sogauged}), (\ref{spgauged})  is 
chiral.  In the first case, (\ref{sogauged}) the $U(N)$ gauge group has an
antisymmetric tensor.  In (\ref{spgauged}) the $U(2M+4)$ group has  a
symmetric tensor. (\ref{sogauged}) and (\ref{spgauged}) are among the 
main results of this paper.

\mysubsection{ The explicit implementation.}

We now turn to explicitly implementing  the basic strategy outlined above. 
It will be important  to consider  the $N=2$ theories with both $SO$ and $SP$ 
gauge groups. 
As was mentioned above, this corresponds to
the two choices of sign in eq.~(\ref{orgroupb}). 
It is  well  known, though not well
understood, from the study of $N=2$ theories 
that  this sign flips in going across an NS brane \cite{EJS}, \cite{LLL},
\cite{B}.  
In our discussion we will, 
correspondingly, allow for the possibility that the twisted-RR
orientifold charge,
eq.~(\ref{orcharge}), also changes on crossing an NS brane.  More
specifically,  we  allow for the  orientifold charges of 
an $SO$ and $SP$ orientifold to 
be different;  for ease of  subsequent  discussion  we  
denote  these integer charges  
by  $-C_{SO}$ and $-C_{SP}$, respectively.  
The corresponding tadpole  cancellation
conditions  eq.~(\ref{orcharge}),  then  are:
\beq
\label{sochrg}
{\rm Tr} ~ \gamma_\alpha \ =\ C_{SO}\ ,
\eeq
and
\beq
\label{spchrg}
{\rm Tr}~ \gamma_\alpha \ =\ C_{SP}\ .
\eeq

We are now ready to explicitly 
implement the truncation procedure outlined above.
We start with Fig.~1,  and  for concreteness first choose the orientifold
charge for the  4-plane lying in between the two NS branes to correspond to  
an $SO$ group.  After discussing this case in some detail we will  then 
turn to  an $SP$ orientifold.   Let the number of D4-branes stretched 
between the NS branes  be $N_1$. 
The  orientifold charge conditions, eqs.~(\ref{sochrg}), (\ref{spchrg}), 
mean that
the configuration we start with 
should  in general  also allow for  semi-infinite D4-branes stretching to the
left and right  of the NS branes (since the O4 plane extends both  to 
the left and right of the 
two NS branes,  semi-infinite D4 branes are needed to 
cancel  its twisted-RR charge).  
We take their number to be  $2 N_L$ and
$2N_R$ respectively.   It is useful  for some purposes  to  think of these 
semi-infinite  4-branes as being  limiting cases of finite ones.
This can be done 
by adding two extra NS branes  one to the  far left and  the other to
the  far right along the $X^6$ direction.   The
full gauge symmetry is  then  $SP(2N_L) \times  SO(N_1) \times SP(2N_R)$
(in particular, this explains why the
number of semi-infinite  4-branes  
stretching to the left  and the right in Fig.~1 must
be even in number).  
Moving the  two  extra NS branes  to infinity turns the $SP(2N_L) \times 
SP(2N_R)$  group into a  global symmetry. 

Strings connecting   4-branes  lying in between the two NS branes  contribute
a  vectormultiplet  of the $SO(N_1)$,  $N=2$ theory. In contrast,  strings 
stretching across   an NS brane  with one end lying on the D4-branes stretched
between the NS branes and the other end on the semi-infinite D4-branes give 
rise to  (half) hypermultiplets which  transform as  
$Q_L ~(\Yfund,\Yfund,1)$  and
$Q_R(1,\Yfund,\Yfund )$  under the $SP(2N_L) \times SO(N_1) \times SP(2N_R)$
symmetry.   It is useful to describe this matter content in terms of the $N=1$
supersymmetry that eventually survives.
The $N=2$ vector multiplet contains an $N=1$ vector multiplet
and a chiral multiplet $\Phi_{SO}$ 
which transforms in the adjoint representation of the gauge group. Each half 
hypermultiplet gives rise to one chiral multiplet. The matter content is then 
given  as: 
\begin{equation}
\label{n2somatter}
\begin{array}{cc|c|c|c}
    &    &SO(N_1)&SP(2N_L)&SP(2N_R)\\ \hline
4-4 & \Phi_{SO} & \Yasymm & 1 &1\\ \hline
left/right& Q_L & \Yfund & \Yfund  &1\\
strings & Q_R & \Yfund & 1&\Yfund 
\end{array} \nonumber
\end{equation}
The theory also has a superpotential 
given by: 
\begin{equation}
\label{n2superpotential}
W ~=~  
{\rm Tr}~ J~ Q_L^T ~ \Phi_{SO} ~ Q_L~+ ~ {\rm Tr}~ J~Q_R^T~\Phi_{SO}~
Q_R~, 
\end{equation}
where $J$ is the $SP$ invariant antisymmetric tensor ($J^2 = -1$). 
This completes our  description of the $N=2$ theory. 

 We now  turn  to identifying the operators
involved in  projecting out states. 
  As mentioned above, the  subgroup of the R symmetries corresponds to a
rotation by $2 \pi/3$ in the $X^4,X^5$ and $X^7,X^8$ planes. This 
means that it acts on the fields as follows:
\begin{eqnarray}
\label{z3infieldtheory}
(A_\mu,~ \lambda)\ &  \rightarrow & \ (A_\mu, ~ \lambda) \nonumber \\
\Phi_{SO}\  &\rightarrow &\ e^{i {2 \pi \over 3}} \ \Phi_{SO} \nonumber \\
Q_{L,R}\ & \rightarrow & \ e^{- i {\pi \over 3}} \ Q_{L,R}~.
\end{eqnarray}

The projection operator also involves a  subgroup of the  gauge symmetry. 
More accurately, in the present context  where there are semi-infinite branes 
present as well,  it involves a subgroup of the $SP(2N_L) \times SO(N_1) \times
SP(2N_R)$ symmetry.  Correspondingly, there are three $\gamma_\alpha$
matrices, eq.~(\ref{egamma}),  $\gamma_\alpha^L, ~ \gamma_\alpha^{SO}, 
~\gamma_\alpha^R$, 
which act on the left semininfinite D4 branes, the 4 branes stretched
between the 
two NS branes, and the semi-infinite 4-branes stretching to the right,
 respectively.  
There is a phase ambiguity, related to these matrices, which we need to 
address
before going further.  The alert reader might have noticed that  in
eq.~(\ref{orgroupa}) 
 we allowed $\gamma_\alpha^3$  to equal a phase and not be necessarily
unity.   The relative phases  of   the $\gamma_\alpha$  matrices 
corresponding  
to  adjacent sets of D4-branes is  physical: it acts  nontrivially 
on the hypermultiplets 
which correspond to  strings going across an NS brane and ending on these
adjacent 4 branes.   By a phase redefinition we can ensure that:
 \beq
\label{phaseso}
(\gamma_\alpha^{SO})^3 =1.
\eeq 
It turns out 
then that\footnote{Evidence in support of this  comes from  
adding six branes placed 
away from the orientifold plane   and in between the two NS branes.
In this case
there are  extra  hypermultiplets,  corresponding to strings
 stretching between the D4-branes and the 6 branes. 
One  is lead to conclude then \cite{GP} that  $(\gamma_\alpha^{(6)})^3=-1$.    
Moving these 6-branes along $X^6$ across an NS brane 
leads  to the creation  of D 4-branes. The  $6-4$ hypermultiplets  then
turn into  the  ones discussed in the text above 
corresponding  to strings stretched
across the NS brane.  This  leads  to the 
conclusion that $(\gamma_\alpha^{L,R})^3 =-1$ as well.}
\beq 
\label{phaselr}
(\gamma_\alpha^{L,R})^3 =-1.
\eeq
  As  we  saw
above,    the semi-infinite D4-branes  can be thought of as limiting
cases of finite
ones  with $SP$ symmetry.   Keeping this in mind,
   our results  for the  respective
phases can be  conveniently summarised as:   
\beq
\label{phasefinal}
(\gamma_\alpha^{SO})^3~=~1~ ,~ (\gamma_\alpha^{SP})^3~=~-1~.
\eeq

We now return to  the main thread of our discussion  regarding  the 
resulting field theory.  For concreteness we   first consider the case where 
$C_{SO},C_{SP} \ge 0$,  eqs.~(\ref{sochrg}), (\ref{spchrg}).
This implies  that (in a suitable basis) :
 \beq
\label{gsoa}
\gamma_\alpha ^{SO}~=~{\rm diag} \{1 \times {\bf 1}_{N+C_{SO}}, 
                                                          ~     \alpha
\times {\bf 1}_N, 
                                                           ~     \alpha^2
\times {\bf 1}_N\} ~,
\eeq 
where ${\bf 1}$ stands for an identity matrix of  appropriate rank. 
We see that  with the above choice,  the number of   D4-branes $N_1$  is 
given by: 
\beq
\label{defn1}
N_1~=  ~3~N~+~C_{SO}~.
\eeq
Similarly,  eqs.~(\ref{spchrg}) and (\ref{phaselr}) imply  that:
\beq
\label{gspl}
\gamma_\alpha ^{L,R}~=~-~{\rm diag} \{1 \times {\bf 1}_{2 M_{L,R}}, 
                                                              ~ \alpha
\times {\bf 1}_{2 M_{L,R}+C_{SP}} 
                           ~    \alpha^2
\times {\bf 1}_{ 2 M_{L,R}+C_{SP}}\} . 
\eeq 
 Eq.~(\ref{gspl}) in turn means  that:
\beq
\label{defnlr}
N_{L,R} ~=~3~ M_{L,R}~ + ~C_{SP}~.
\eeq

Let us now consider the effect of the projection operators on the various 
fields in the theory.  As we saw above the gauge boson is invariant under 
the $\IZ_3$ R symmetry.  The projection operator  acts on it by:
\beq
\label{projgauge}
A_{\mu} ~\rightarrow  ~\gamma_\alpha^{SO} ~ A_{\mu}~
(\gamma_\alpha^{SO})^{-1}.
\eeq
Eq.~(\ref{gsoa}) shows then that  the final theory has a  $SO(N+C_{SO}) \times
U(N)$ gauge symmetry\footnote{The $U(1)$ subgroup of the $U(N)$ 
will turn out to be  anomalous;
we will return to this later on. }. 
Similarly the $SP(2N_L)$ and $SP(2N_R)$  global symmetries are  broken to
$U(2M_L+C_{SP}) \times SP(2M_L)$ and   $U(2M_R+C_{SP}) \times SP(2M_R)$
respectively.  The matter content   can be found in a similar manner by  asking
which states are invariant under  the combined action of the $\IZ_3$
R symmetry,
$\gamma_\alpha^{SO}$ , $\gamma_\alpha^L$, and $\gamma_\alpha^R$. 
It can be read off from  eq.~(\ref{sogauged}), after  making the  changes: 
$U(2M_{L,R}+4) \rightarrow U(2M_{L,R}+C_{SP})$,  $SO(N+4) \rightarrow
SO(N+C_{SO})$.

We are now ready to impose the constraint of anomaly cancellation.  Demanding
that  the $SU(N)$ subgroup of the $U(N)$ gauge symmetry is  anomaly free
gives  the condition:
\beq
\label{soano}
C_{SO}~-~2~C_{SP}~=~- ~4~.
\eeq
 
So far, we have considered the case where the orientifold charge (between
the NS branes)
was  chosen to correspond (in the $N=2$ case) to  an $SO(N)$ gauge theory. 
Now we can consider the case where this  is  an $SP$ gauge  symmetry instead,
restricting ourselves again to the case $C_{SO},C_{SP} \ge 0$.
In this case  a similar analysis reveals that the final gauge symmetry is 
$SP(2M) \times U(2M+C_{SP})$.  The global symmetry is
correspondingly given by $SO(N_L+C_{SO}) \times U(N_L) \times
SO(N_R+C_{SO}) \times U(N_R)$. The  matter content,  which survives the
projection is given by  eq.~(\ref{spgauged}) after making the changes
$SO(N_{L,R}+4) \rightarrow SO(N_{L,R}+C_{SO})$ and $U(2M+4)
\rightarrow U(2M +C_{SP}) $. 

Anomaly cancellation for the $SU(2M+C_{SP})$ now implies that 
\beq
\label{anosp}
2~C_{SO} ~-~C_{SP}~=~4~.
\eeq
Eqs.~(\ref{soano}) and (\ref{anosp}) now show that 
\beq
\label{anosol}
C_{SO}~=~C_{SP}~=~4~.
\eeq

So far we have restricted ourselves to the case $C_{SO},C_{SP} \ge 0$. 
On considering the other possibilities one finds no  solutions consistent
with the requirements of anomaly cancellation.  Thus  one is lead to the unique
values of the orientifold charges, 
eq.~(\ref{anosol} )\footnote{While we do not do so here, 
one can show that the solution, eq.~(\ref{anosol}), is consistent 
with the discrete subgroup of the R symmetry, eq.~(\ref{z3infieldtheory}),
being non-anomalous as well.}. 
 
To summarize,  for an $SO$ orientifold stretched in between the two NS
branes  the gauge group  is  $SO(N+4) \times U(N)$, the global
symmetry group is  $SP(2M_L) \times  U(2M_L+4) \times  SP(2M_R)
\times U(2M_R+4)$. The matter content is as given in  eq.~(\ref{sogauged}) 
Note that the theory is chiral and contains  and antisymmetric tensor
of the $U(N)$ gauge group.  The theory inherits a superpotential from   
eq.~(\ref{n2superpotential}): 
\begin{equation}
\label{sogaugedsuperpotential}
W ~= ~\sum_{i = L, R}  ~ A \cdot \bar{q_i} \cdot \bar{q_i}  ~+~ 
      \bar{Q} \cdot v_i \cdot q_i~.
 \end{equation}

For  the  $SP$ orientifold the gauge symmetry is 
$SP(2M) \times U(2M+4)$,  the global symmetry is $SO(N_L+4) \times U(N_L)
\times SO(N_R+4) \times U(N_R)$. 
  The matter content is shown in 
eq.~(\ref{spgauged}).  The corresponding superpotential is: 
\begin{equation}
\label{spgaugedsuperpotential}
W ~= ~\sum_{i = L, R}  ~ S \cdot v_i  \cdot v_i   ~+~ 
       Q  \cdot q_i \cdot \bar{q}_i
 \end{equation}
We note that to prevent a global anomaly in the $SP(2M)$ gauge symmetry
$N_{L}$ and $N_R$ must both be either even or odd. 

One final comment is in order before we end this section.  
As was  mentioned above 
in the $SO$ case, the $U(1)$ subgroup of $U(N)$  is anomalous.  
Similarly, in the $SP$ case the $U(1)$ subgroup of $U(2M+4)$  is anomalous.
 The role of   anomalous  $U(1)$s   in  D-brane theories  
was discussed in ref.~\cite{DM}, 
where they  were shown to  be  important  in studying the low energy
dynamics.  We  will not discuss this issue in great detail.  The important 
point is
that the $U(1)$s are in fact broken. The anomalies are cancelled by the Green
Schwarz mechanism;  the axion fields involved  come from the twisted RR fields,
ref.~\cite{DM}.  
The one feature that will be important in the present discussion is that 
each $U(1)$ gives a $D$-term contribution to the potential energy, which is 
important in determining the moduli space of the theory.   The  axions
mentioned above  have  partners whose expectation value  corresponds to 
Fayet-Iliopoulos $D$ terms for the anomalous $U(1)$s.  Geometrically these 
expectation values correspond  to blowing up  
the orientifold background\footnote{In fact,  in the 
orientifold case, eqs.~(\ref{sogauged}), 
(\ref{spgauged}),
since the anomalous $U(1)$s are not traceless, a contribution to
the FI terms is 
generated at one loop as well. }.  We will have more to say about blowing up 
 orbifolds and orientifolds in  Section  5. 

\mysection{ The Classical Moduli Space.}

We now turn to comparing the classical moduli space of the field theories
discussed above with the allowed motions for the brane configuration.  Detailed
agreement between the two 
provides additional evidence that the correct 
field theories have been identified.  

For simplicity  we keep the number of semi-infinite 4-branes  in Fig.~1 to be 
the minimum  required by the tadpole conditions, eq.~(\ref{sochrg}), 
(\ref{spchrg}), with eq.~(\ref{anosol}).  This means   
$M_{L,R}=0$ in eq.~(\ref{sogauged}) and the $\bar{q}_{L,R}$ 
fields are absent in the $SO$ case; similarly, 
 $N_{L,R}=0$ in eq.~(\ref{spchrg}), with both $q_{L,R}$ and $\bar{q}_{L,R}$ 
being absent in the 
SP case.   The  generalization
for the non-minimal case is straightforward and does not add anything new.

\mysubsection{The $SO$ theories.}

There are two cases to be discussed, 
corresponding to the $SO$ theories,
eq.~(\ref{sogauged}), and  the $SP$ theories, eq.~(\ref{spgauged}).  We first 
consider the $SO$ theories.   As follows from eq.~(\ref{sogauged}),
a general configuration has $3N+4$ branes  coincident  with  the
orientifold plane. Starting with this  configuration, branes can be 
moved 
away from  the orientifold plane along the $X^4,X^5$ directions.   Each brane,
away
from  the orientifold  plane, has   five  images under the  $\IZ_6$
orientifold group. 
This means that, counting  images,
 we can move   sets of six branes away from the
orientifold point.   Moving one set of six branes should correspond to
Higgsing the gauge symmetry $SO(N+4) \times U(N) \rightarrow SO(N+2) \times
U(N-2)$.  Since the branes move in $X^4,X^5$ there is one complex modulus
associated with this motion.  

It is useful in the subsequent discussion to distinguish between two cases. 
If  the total number of branes we started
with is  even, i.e. $N$ is even,  
this process can be carried on till  $3N$ branes have
been moved away,
 leaving  four branes stuck at the orientifold.  
The  four branes saturate the tadpole condition, 
Tr$(\gamma_\alpha^{SO})=4$, eqs.~(\ref{sochrg}),
(\ref{anosol}),  and give 
rise to an $SO(4)$ gauge symmetry.
The branes that have been moved away  
contribute a  $U(1)^{N\over 2}$ factor,
leading to a  full $SO(4) \times U(1)^{N\over 2}$ gauge symmetry. 
There are $N/2$ moduli parametrizing the allowed motions of the 4-branes. 

 If   $N$ is odd,
a maximum of $3(N-1)$ branes  can be moved away,
   leaving  seven branes
stuck at  the orientifold.  These branes generically should give rise to  an 
$SO(4)$  gauge group, 
leading to a full $SO(4) \times  U(1)^{(N-1) \over 2}$ 
gauge symmetry\footnote{Eq.~(\ref{sogauged}) would  
lead us to believe that the gauge group  for
the branes left  at the orientifold  is $SO(5) \times U(1)$.  However, the 
$U(1)$ factor is anomalous and broken; generically it also has  a non-zero 
FI term. To meet the D-flatness condition for 
the $U(1)$ when the FI term is non-zero,  the 
$\bar{Q}$ field must acquire a vev breaking 
$SO(5) \rightarrow SO(4)$ (see discussion after eqns.~(\ref{soflatQodd}),
(\ref{soflatAodd})~).}. 
In this case there are $(N-1)/2$ moduli 
parametrising the  set of allowed motions. Finally, note that in both the even 
and odd cases case when 
$N_1$ branes  come together away from the orientifold plane (as do their
images)
the  corresponding $U(1)^{N_1}$  factor in the gauge symmetry is enhanced to 
$U(N_1)$.

Let us see if the analysis in the field theory  agrees with this. 
Note that of the field content in eq.~(\ref{sogauged}), only the $\bar{Q}$
and $A$
fields  arise from the adjoint $\Phi_{SO}$, eq.~(\ref{n2somatter}). 
The other fields  originate from
the (half) hypermultiplets $Q_{L,R}$.  For the $N=2$ supersymmetric theory  it
is well known that  the brane construction   with semi-infinite 
4-branes   describes  the Coulomb branch  of the $N=2$ theory
along which  $\Phi_{SO}$ 
gets a vev but not the hypermultiplets $Q_{L,R}$ \cite{LLL}, \cite{B}.  
Accordingly,  one expects   for the $N=1$
theories being discussed here  that the allowed  brane  motion   will 
correspond to  the branch of the moduli space  where  only  $\bar{Q}$ and $A$
get a  vacuum expectation value.  Below, we verify this. 

For the $N$ even case the solutions to the $SO(N+4)$ 
and $SU(N)$ D-flatness conditions are: 
\begin{eqnarray}
\label{soflatQ}
\bar{Q} ~=~ \left( \begin{array}{ccccccccc}
v_1 & 0& \ldots&0 &0 & 0 &0 & 0 &0 \\
0 & v_1& \ldots&0 & 0 & 0 &0 & 0 &0 \\
0 & 0 & \ldots & 0 & 0 & 0 &0 & 0 &0 \\
\ldots &  \ldots&  \ldots&   \ldots&   \ldots&  \ldots & \ldots &  \ldots & 
\ldots  \\
0 & 0 & \ldots & v_{N/2} & 0& 0 &0 & 0 &0  \\
0 & 0 & \ldots & 0 & v_{N/2} & 0 &0 & 0 &0  \end{array} \right)~,
\end{eqnarray}
\begin{equation}
\label{soflatA}
A~=~ {\rm diag}~\left( ~a_1 ~ \sigma_2 , ~a_2 ~\sigma_2, \ldots
, ~a_{N/2} ~ \sigma_2 ~\right)~, ~ {\rm with} ~
|v_i|^2 ~ = ~2 ~|a_i|^2 ~+ ~c ~.
\end{equation}
Since these are the only fields that get expectation values,
the $F$-flatness
conditions are automatically met, see eq.~(\ref{sogaugedsuperpotential}). The 
constant $c$ in eq.~(\ref{soflatA}) is fixed, by the anomalous $U(1)$ D-flat 
condition,
to be proportional to its  Fayet-Iliopoulos term. This is easy to see upon
substituting the solutions (\ref{soflatQ}), (\ref{soflatA}) into the
anomalous  U(1) D-term equation,
$2 {\rm Tr} A^\dagger A - {\rm Tr} \bar{Q} \bar{Q}^\dagger = \xi_{FI}$.
Clearly,   the flat directions
(\ref{soflatQ}) and (\ref{soflatA}) preserve a diagonal 
$SO(2)^{N/2} = U(1)^{N/2}$ abelian gauge symmetry,
times a nonabelian $SO(4)$, exactly
as predicted by the brane picture.

We also note that along these  flat directions the superpotential, 
eq.~(\ref{sogaugedsuperpotential}), 
 gives mass 
to all the $q_{L,R}$ fields  and to the $SO(N)$ components of 
the $v_{L,R}$ fields.
This can be understood in the brane picture as well:
when the D4-branes moves away from the orientifold plane,
the open strings 
connecting them to the adjacent semi-infinite 4-branes get stretched. 

The $SO(4)$ group left unbroken has 
eight vectors and is non-asymptotically free. Consequently, 
in the quantum theory, we expect
the $SO(4)$ gauge coupling to go to zero in the infra-red and for the massless
spectrum to include the $SO(4)$ gauge bosons as well as the eight vector 
matter fields.

For the odd  $N$ case  a similar field theory analysis for the flat directions 
yields for
the
field $\bar{Q}$:
\begin{equation}
\label{soflatQodd}
\bar{Q} ~=~ \left( \begin{array}{cccccccccc}
v_1 & 0& \ldots&0 &0 & 0 &0 & 0 &0 &0\\
0 & v_1& \ldots&0 & 0 & 0 &0 & 0 &0 &0 \\
0 & 0 & \ldots & 0 & 0 & 0 &0 & 0 &0 &0\\
\ldots &  \ldots&  \ldots&   \ldots&   \ldots &\ldots&  \ldots & \ldots & 
\ldots & \ldots  \\
0 & 0 & \ldots & v_{N-1\over 2} & 0& 0 &0&0 & 0 &0  \\
0 & 0 & \ldots & 0 & v_{N-1 \over 2} & 0&0 &0 & 0 &0 \\
 0 & 0 & \ldots & 0 &0& \sqrt{c} &0 &0 & 0 &0 \end{array} \right)~,
\end{equation}
while the antisymmetric tensor now has rank $N-1$:
\begin{equation}
\label{soflatAodd}
A~=~ {\rm diag}~\left( ~a_1 ~ \sigma_2 , ~a_2 ~\sigma_2, \ldots ,
~a_{N-1 \over 2} ~ \sigma_2,~ 0 ~\right) ~, ~ {\rm with} ~
|v_i|^2 ~ = ~2 ~|a_i|^2 ~+ ~c ~.\end{equation}
The constant $c$ for the odd case is also fixed to
be proportional to the anomalous $U(1)$ Fayet-Iliopoulos term. Note that 
if the full FI term (sum of tree level and one loop) is non-zero,
$c \ne 0$, and  $SO(5)$ is broken down to $SO(4)$. Thus, in the odd-$N$ case 
generically the unbroken gauge symmetry is   expected to be
$U(1)^{N-1 \over 2}\times SO(4)$. The superpotential gives a mass to 
the $q_{L,R}$ fields and the $SO(N)$ components of the $v_{L,R}$.

The Coulomb branch described above in eqns.~(\ref{soflatQ}, \ref{soflatA})
(or (\ref{soflatQodd}, \ref{soflatAodd}) for the odd-$N$ case) can
also be given a gauge invariant description: from the fields $\bar{Q}$ and $A$
we can form the $U(N)$ invariant $X = A \cdot \bar{Q} \cdot \bar{Q}$, which
transforms as an adjoint of $SO(N + 4)$. The rank of $X$, however, is at most
$N$ (or $N-1$, if $N$ is odd). Therefore, the independent $SO(N+4) \times U(N)$
invariants that describe the Coulomb branch are Tr$X^{2 j}$,
with $j = 1,...,[N/2]$. These correspond to the positions of
the $[N/2]$ physical D4-branes away from $v=0$.
 Note also that the
$SO(N+4)\times SU(N)$ invariant ``baryon"
 $\bar{Q}^{2 N}$ is not allowed by the anomalous
$U(1)$ (this corresponds to eliminating $c$ from (\ref{soflatA}) by the U(1)
D-term). The pattern of symmetry breaking is, of course, the same as the one
described above via the explicit solutions along the flat directions.

We end this section with one final  comment. Above, we found that the unbroken 
gauge symmetry is generically, $U(1)^{N-1 \over2} \times SO(4)$. As in the 
even case, the $SO(4)$ has eight matter fields in the vector representation
and is non-asymptotically free. Once again, in the quantum theory,
 we expect the $SO(4)$ gauge bosons and the eight matter fields to be present
in the massless spectrum. 

\mysubsection{The SP theories.}

The description for the $SP$ case is similar to the $SO$ case and 
we only present it briefly. As in the $SO$ case we expect that the
allowed brane 
motion should correspond to the branch of moduli space where 
only the descendants of the $N=2$ adjoint, $Q, \bar{S}$, 
eq.~(\ref{spgauged}), get an expectation value. 
 In this case the tadpole condition, eq.~(\ref{spchrg}) and eq.~(\ref{anosol}),
 requires
$6M+8$ branes to be coincident with the orientifold plane. $6M$ of these 
can be moved away from the orientifold plane; this motion is parametrized 
by $M$ moduli. The remaining eight branes are stuck at the orientifold 
plane and saturate the tadpole condition, Tr$(\gamma_\alpha^{SP})=4$ 
(with $\gamma_\alpha^{SP}$ of the form given by eq.~(\ref{gspl})). These eight 
branes give rise 
to an $SU(4)$ gauge symmetry, while the branes away from the orientifold
plane contribute a $U(1)^M$ factor leading to a full $SU(4) \times U(1)^M$
gauge group. In field theory the moduli space is parametrised by the $M$ gauge 
invariants Tr $(S \cdot Q \cdot Q)^{2j}$, with $j = 1,...,M$. The gauge 
group is $SU(4) \times U(1)^M$ and again agrees with  what was expected 
from brane considerations. Also, 
along these directions the $U(2M)$ components of the $v_{L,R}$ fields get 
heavy eq.~(\ref{spgaugedsuperpotential}).

We saw above that a $SU(4)$ subgroup is left unbroken. It 
has a symmetric tensor and eight antifundamentals as matter fields
and is asymptotically free. One would like to understand the quantum behavior
of this theory, in particular whether it confines or not and how it 
global symmetries are realised. Unfortunately, these turn out to be difficult 
questions and we will have to leave them unanswered. What complicates
the analysis is  the fact that the theory has no flat directions.
The superpotential, 
eq.(\ref{spgaugedsuperpotential}),  lifts all 
flat directions involving the $v_i$ fields leaving only one flat direction
along which the field $S$ gets a vev breaking $SU(4) \rightarrow SO(4)$. 
This direction in turn is lifted by the D term of the anomalous U(1). 
We can say one important thing about the ground state of this theory:
it does not break supersymmetry. 
One might have worried for example that along the flat direction, mentioned 
above, where $S$ gets a vev breaking $SU(4) \rightarrow SO(4)$,
gaugino condensation in the $SO(4)$ generates a superpotential that pushes
the $S$ to
large vevs, in conflict with the $U(1)$ $D$-term requirement and leading to
supersymmetry breaking. However,
since  $SO(4) \simeq SU(2) \times SU(2)$, there is a branch along which the
gaugino condensates from the
two $SU(2)$s contribute oppositely  and cancel;
 along this branch the quantum theory has a supersymmetric vacuum.

Finally, the field theory analysis above both for the $SO$ and $SP$ cases
 was restricted to generic points in the moduli space. On extending it to
points of enhanced gauge symmetry one finds again agreement with the
expectations from the brane picture.

\mysection{The Hyperelliptic Curves from M Theory.}
 
So far in our considerations we  have considered the brane configuration
in the 
Type $II A$ limit.  We now turn to considering it in M theory;
this will allow us 
to determine the spectral curves for the Coulomb phase described in section 3. 
Our  analysis  is   based on the important observation of ref.~\cite{wittenone}
that in M theory,
the brane configuration of Fig. 1 can be thought of as 
a single   NS 5-brane which is smooth  on the eleven
dimensional Planck scale.  
In ref.~\cite{wittenone},
 the M theory 5-brane world volume had infinite 
extent in the $X^0, X^1, X^2,
X^3$ coordinates,
while spanning a two dimensional surface in a four manifold, 
parametrized by $v=X^4+iX^5$ and $t={\rm exp}(-(X^6+iX^{10})/R)$.  
Refs.~\cite{LLL} and \cite{B} extended
this study to the $SO$ and $SP$ cases 
by introducing  an orientifold  4-plane.  The
effects of the orientifold plane were incorporated by working in the  covering 
space, $X^4,X^5,X^7,X^8,X^9$, and restricting to configurations symmetric
under  
\beq
\label{orv}
(X^4, X^5, X^7, X^8, X^9 ) \rightarrow - (X^4, X^5, X^7, X^8, X^9 ).
\eeq

For the case at hand,  we consider placing the brane configuration of 
Fig.1 in  a $\IC^2/{\IZ_3}$ orbifold background, 
given by identifying points as in eq.~(\ref{defzthree}). A more 
convenient representation 
of this   orbifold is obtained by embedding it as a hypersurface in $\IC^3$:
\beq
\label{fancydef}
yz-x^3=0.
\eeq
The coordinate mapping is $y=v^3,z=w^3, x=vw$.  The 5-brane is described
as a curve $\Sigma$ in $C^3 \times R^1 \times S^1$. The  effects of the 
orientifold 4-plane  are  incorporated, as in the $N=2$ case, by restricting to
surfaces  symmetric under  $(y,w )\rightarrow -(y,w) $.   The surface
$\Sigma$ is smooth except at the orientifold point  $y=z=x=0$ and can be 
parametrized by $y$ and $t$ with $z$ set equal to zero.  

There is one  important limitation in  our  discussion below that  should be 
mentioned at the outset. We saw above that the $SO$ case gives rise to a $SO(4)
\times U(1)^{[N] \over 2}$ gauge symmetry; the $SO(4)$ group is infra-red
free and we expect its ultra-violet degrees of freedom to be present in 
the infra-red as well. 
The $SP$ case, on the other hand,  gives rise  to a $SU(4) \times U(1)^M$
unbroken gauge symmetry; here the inra-red behavior of the 
$SU(4)$ theory is difficult to analyse since it  is asymptotically free
and has no moduli. The
discussion  below will lead to the spectral curves from which the 
gauge couplings etc. of the $U(1)$ factors can be found, but we will not
address  how the physics of the $SO(4)$ or $SU(4)$ gauge theory  can be
deduced from M theory considerations. We expect this to be  a fairly
non-trivial question. In the Type IIA limit, the $SO(4)$ ($SU(4)$) symmetry
arose because of 4-branes stuck to the orientifold. Understanding this
in M theory probably  involves
blowing up the orbifold and working in the
resulting ALE space (see the discussion in section 5 in this regard), in
particular the  non-trivial two cycles of the ALE space should play an
important
role in this. We hope to return to this question in the future.

In the analysis below, our
basic strategy  is   to
start with the $N=2$ curves expressed in terms of the coordinates  $t,~v$.  
Incorporating the   $\IZ_3$ orbifold   results in setting  some  of the order
parameters of the $N=2$ curves to zero.  Then, 
 re-expressing the curves  in terms of $t$ and $y$ gives rise to the 
final form of curves.   The $U(1)$ gauge fields, in the resulting 
world volume theories,   
correspond to  harmonic
differentials  odd   under  the $y \rightarrow -y$ symmetry and  we  see below
that their   number
 agrees with the field theory analysis, section 3.     
Some  additional checks on the curves are carried out as well  and  shown 
to  match  field  theory expectations.  There are three classes of $N=2$ 
theories with even $SO$, odd $SO$, and $SP$ gauge symmetries,
which give rise to three classes of  $N=1$ theories.  We consider these in
 turn.

\mysubsection{The even $SO$ case.}

To begin, we consider the $SO(2n+4) \times U(2n)$ gauge theory, eq.
(\ref{sogauged}), with 
$M_L=M_R=0$ (incorporating additional semi-infinite four branes is trivial and
does not add anything to the discussion; see eq.~(\ref{fcmeven})).
As discussed in section 2,  the  corresponding $N=2$ theory  has 
$6n+4$ branes placed  between the two NS branes and eight semi-infinite branes
extending to the  left and right respectively.  
The corresponding curve is given by \cite{LLL}, \cite{B}:
\begin{equation} 
\label{n2soevencurve} 
v^{10}~t^2~+~t~\left(~ (v^2)^{3n+2} ~+ ~u_2
~(v^2)^{3n+1} ~+~ \cdots u_{6n+4} ~\right)  ~+ ~v^{10}~=~0~.
\end{equation}
Imposing the $\IZ_3$ symmetry sets several terms above to zero and gives rise
to the curve: 
\beq
\label{incurve}
v^{10}~t^2~+~t~\left(~(v^2)^{3n+2} ~+~ {\tilde u}_1~(v^2)^{3n-1} ~+~  
\cdots +{ \tilde u}_n~(v^2)^{2} \right)~ +~v^{10}~ =~0~.
\eeq
We see that  the curve above can be written as the product of two factors:
\beq
\label{faccurve}
v^4~ \left[~v^6~t^2~ +~ t~\left(~(v^2)^{3n} ~+~
{\tilde u}_1~(v^2)^{3n-3}~ +~ \cdots ~+~ {\tilde
u}_n \right)~ +~ v^6 \right]~ =~0~.
\eeq
The analysis  in  section 3.1 leads us to expect that $SO(2n+4) \times
U(2n) \rightarrow SO(4) \times U(1)^n$. The discussion below shows that 
the second component in eq.(\ref{faccurve})  corresponds to the
spectral curve from which the gauge couplings etc. of the $U(1)^n$
group can be calculated. We expect the  $v^4$ factor to be important in
understanding the $SO(4)$ unbroken symmetry, but as mentioned above we will 
not pursue this any further here.  

 Discarding the $v^4$ factor and  substituting, $v^3 \equiv y$ as
mentioned above then  gives the final form of the curve:
\beq
\label{fceven}
y^2~t^2 ~+~ t~\left(~(y^2)^{n} ~+~ {\tilde u}_1~(y^2)^{n-1}~ +~ \cdots~ +~
{\tilde u}_n~ \right)~ +~y^2 ~=~0~.
\eeq
Eq.~(\ref{fceven}) meets a number of checks, as we now discuss. 

First,  it  has $n$ order parameters, ${\tilde u}_j, j=1,\cdots,n$. This 
agrees with the field theory analysis in section 3.   
The moduli ${\tilde u}_j$  can be related, semiclassically,
to the projections of the $N=2$ moduli that are $\IZ_3$ invariant:
${\rm Tr} (\Phi_{SO}^2)^{3 j}$, $j = 1,...,n$. In the $N = 1$ theory,
the $N=2$ adjoint decomposes into the fields $\bar{Q}$ and $A$, and using 
this identification, we can conclude that, 
semiclassically, the moduli of the $N=1$ theory are
${\rm Tr} (A \bar{Q}^2)^{2 j}$, in accord with the field theory expectations.

Second, as  mentioned above, 
 the $U(1)$ gauge bosons correspond
to  holomorphic differentials that are odd under 
$y \rightarrow -y$. One can show 
from eq.~(\ref{fceven}) that these are $n$ in number, again in accord with the 
field theory analysis.  

Third, the curve, eq.~(\ref{fceven}), is consistent with all the symmetries 
(exact and anomalous) of the field theory.  To see this  
it is convenient to  introduce in  eq.~(\ref{fceven})
 an explicit dependence  on  the strong coupling scales of 
the $SU(2n)$ and
$SO(2n+4)$ groups.   After  rescaling $t$ and $y$,  dimensional analysis and 
the requirement of  invariance with respect to  the $U(1)$ (anomalous)
subgroup of  the $U(2n)$  gauge symmetry fixes  the curve to be:
\beq
\label{fcseven}
y^2~t^2~ +~ t~\left(~(y^2)^{n}~ +~ {\tilde u}_1~(y^2)^{n-1} ~+ ~\cdots 
{ \tilde u}_n~ \right)~ +~
\Lambda_{SU}^{2 b_{0}^{SU}}
~\Lambda_{SO}^{b_0^{SO}}  \ y^2 ~=~0~.
\eeq
One can now verify that eq.~(\ref{fcseven}) transforms appropriately under 
all the other symmetries of the  field theory as well.
(In eq. (\ref{fcseven}),
 $b_0^{SO(2n+4)} = 4n - 2$  and $b_0^{SU(2n)} = 4n - 5 $ are 
the one loop beta
function coefficients of the $SO(2n+4) \times SU(2n)$ theory, 
with $M_L=M_R=0$, respectively
 (\ref{sogauged})\footnote{We note that,  since $2 b_0^{SU(2n)} + b_0^{SO(2n+4)}
= b_{0, N=2}^{SO(6n + 4)}$,  the total power of the scale dependence in 
eq.~(\ref{fcseven}), $\Lambda_{SU}^{2 b_{0}^{SU}}
\Lambda_{SO}^{b_0^{SO}}$ ,  matches that in  the $N=2$ theory, 
$\Lambda^{ b_{0, N=2}^{SO(6n +4)}}$.}).

Finally, one can consider  submanifolds   of  the moduli space that correspond
classically to an enhanced gauge symmetry and ask if the curve reduces  there 
to the expected form. As an example, consider  a situation where  $n$ physical 
branes come close together far from the orientifold, resulting in  an $U(1)^n
\rightarrow U(n)$ enhanced symmetry.  The   $U(N)$   theory  has  an adjoint
field and we expect the curve to reduce to that for  an $N=2$ $SU(N)$
theory\footnote{ The overall $U(1)$ is not relevant for this discussion.} along
this flat direction.  To see that this is indeed the case,
rewrite the curve in the 
form:
\begin{equation}
\label{flatcheck}
y^2~t^2 ~+~ t ~( y^2 -a_1^2)~(y^2 -a_2^2)~\ldots~ 
(y^2 - a_n^2) ~+~
 ~\Lambda_{SU}^{2 b_0^{SU}}
 ~ \Lambda_{SO}^{b_0^{SO}}~y^2 ~= ~ 0~.
\end{equation}
The parameters $a_i$ in (\ref{flatcheck}) correspond to the positions of the 
$n$ physical D4-branes in the $y$ plane. The configuration with enhanced
$U(n)$ we  described above corresponds to 
$a_i = a + \delta a_i$, with $\sum_{i = 1}^n \delta a_i = 0$; $a$ describes
the center of mass motion of the $n$ physical branes in the $y$ plane, whereas
the $\delta a_i$ are the small fluctuations around the center of mass. In the
vicinity of $y = a$, i.e. for 
$y = a + \delta y$, the curve (\ref{flatcheck}) becomes, keeping only the 
leading terms in $\delta y, \delta a_i$:
\begin{equation}
\label{flatcheck1}
a^2~ t^2 ~+~ t~ ( 2 ~a )^n ~(\delta y - \delta a_1)~ (\delta y -
\delta a_2) ~\ldots ~ (\delta y - \delta a_n) ~+~
  ~\Lambda_{SU}^{2 b_0^{SU}}~ 
 \Lambda_{SO}^{b_0^{SO}}~a^2 ~= ~ 0~.
\end{equation}
But this is precisely the curve describing  the  $SU(n)$ theory
with an adjoint,
with the correct strong coupling scale,  
as can be seen by rescaling 
away the coefficients in front of $t^2$ and $t$, and remembering that 
$\sum_{i = 1}^n  \delta a_i = 0$.

We end  the discussion for the $SO(2n)$ case with two comments.
First, from eq.~(\ref{fceven}),
 we  see that  (for ${\tilde u}_n \ne 0$ ) as $y \rightarrow 0$,
$t \rightarrow 0$ and $\infty$. Thus the orientifold plane pushes the
NS 5-brane 
out to infinity in the small $y$ region;
this is analogous to what happens in the $SO(2n)$ $N=2$ case.  
Second, above we considered the case with the minimal number of 
semi-infinite 4-branes, i.e. $M_L=M_R=0$.  In the more general case where
$M_L,M_R \ne 0$ the curve can be written down in an analogous fashion. It is
given by:
\beq
\label{fcmeven}
y^2 ~t^2~\prod_{i=1}^{M_L}~(y^2-a_i^2) + e~t~\sum_{j=0}^{n}~
                            {\tilde u}_i~(y^2)^{n-i}  + f~y^2 
\prod_{k=1}^{M_R}~(y^2-b_k^2)
                                       =0.
\eeq

\mysubsection{The odd $SO$ case.}

Our discussion  here will closely parallel   the previous section,
consequently we   omit  many of the details and only   mention  the important
points.  To begin,  consider  the  theory with  $SO(2n+5) \times U(2n+1)$
gauge symmetry  and $M_{L,R}=0$, eq.~(\ref{sogauged}). The 
corresponding $N=2$ theory is obtained
by  taking a configuration containing   $6n+7$  4-branes between the two
NS branes and eight semi-infinite 4-branes stretching to the left and right
respectively.  The $N=2$ curve is :
\beq
\label{codd}
v^9~t^2~+t~\left(~(v^2)^{3n+3}~ +~ u_2~(v^2)^{3n+1} ~+~ \cdots 
~+ ~u_{6n+6} \right)~ +~ v^9~=~0~.
\eeq
Notice that this curve is invariant under the combined transformation,
$v \rightarrow -v $, $t \rightarrow -t$. 
After imposing the $\IZ_3$ symmetry  to set various terms to zero and 
substituting $y \equiv v^3 $ we get:  
\beq
\label{cintod}
y^3~t^2~+~t~\left(~(y^2)^{n+1} ~+~ {\tilde u}_1~(y^2)^{n}~ +~ \cdots~ +~
 {\tilde u}_n~y^2~ +~ 
{\tilde u}_{n+1}~\right) ~+~y^3~=~0~ .
\eeq 
It would  seem that the curve has $n+1$ order parameters, 
${\tilde u}_j,~ j=1, \cdots, n+1$; however, a semiclassical  analysis  shows 
that  ${\tilde u}_{n+1}$ is not an order parameter.  

Let us pause briefly to discuss this.  Note that  semiclassically,
this parameter is the descendant of  $u_{6n+6}$ in the $N=2$ theory, 
eq.~(\ref{fcodd}), which  is proportional to the product of the squares of the
$3 n + 3$ nonzero eigenvalues of $\Phi_{SO(6n+7)}$ (recall that the eigenvalues
come in pairs). The decomposition of the adjoint
of $SO(6n+7)$ in terms of the fields $\bar{Q}, A$ has the form (in  a basis
where the $SO$ adjoint is an antisymmetric matrix):
\begin{equation}
\label{decomposeadj}
\Phi_{SO(6n+7)} ~=~ \left( \begin{array}{ccc}
0 &  \bar{Q}^T  & - i \bar{Q}^T  \\
- \bar{Q} & A & i A \\
i \bar{Q} & i A & - A \end{array} \right) ~.
\end{equation}
Then, substituting the solutions for the flat directions (\ref{soflatQodd},
\ref{soflatAodd}) into (\ref{decomposeadj}),  we see that  
$\Phi_{SO(6n+7)}$ can at most have $3 n + 1$ 
nonvanishing pairs of eigenvalues
along the flat directions. 
Hence, semiclassically, $\tilde{u}_{n+1} =0$. This  shows that in the full
quantum theory $\tilde{u}_{n+1}$ cannot be continuously varied but leaves  
open the possibility that  it is a constant different from   zero (and dependent
on the two strong coupling scales).   
At  present, we do not know how to determine this constant  from a first
principles analysis.   We will see below though that
setting   $\tilde{u}_{n+1} =0$ will  yield a curve that meets several checks. 

 Putting  $\tilde{u}_{n+1} =0$, in eq. (\ref{cintod}) gives a
curve which factorizes. 
The analysis in section 3.1  shows  that  generically , along the branch of
moduli space under discussion here, $SO(2n+5) \times U(2n+1)  \rightarrow SO(4)
\times  U(1)^n$.  In analogy with the even $SO$  curve we expect
that the spectral
curve for the $U(1)^n$ subgroup is given by  discarding the overall
factor of $y^2$. This
\footnote{Once again the  $y^2$ factor should be important in understanding the
$SO(4)$ symmetry in M theory.} yields the  final form of the curve: 
\beq
\label{fcodd}
y~t^2~+~t~\left(~ (y^2)^n
~ +~ {\tilde u}_1~(y^2)^{n-1} ~+~ \cdots~ +~{\tilde u}_n \right)~ +~y~=~0~.
\eeq 

Let us describe  some of the checks  that  the curve,  eq.(\ref{fcodd}), meets.
First, it has $n$ moduli.  In fact,
from eq.~(\ref{decomposeadj}),  it follows  that semiclassically  these
correspond to  Tr$(A {\bar Q} \bar{Q} )^{2j}$, $j=1 \cdots, n$.  
Second, it gives rise to   $n$ holomorphic differentials odd under 
$y \rightarrow -y, t \rightarrow -t$.  This agrees with the $n$ $U(1)$
gauge bosons expected from  the field theory  analysis. 
Finally,  after rescaling $t$ and $y$ above the curve can be written
in the form:
\beq
\label{fcsodd}
y~t^2~+~t~\left(~(y^2)^{n} ~+~ {\tilde u}_1~(y^2)^{n-1} ~+~ \cdots~ +~ 
{\tilde u}_n \right)~ +~y
 ~\Lambda_{SU}^{2 b_0^{SU}}
 ~ \Lambda_{SO}^{b_0^{SO}}~= ~ 0, 
\eeq
which is easily seen to be consistent with all the symmetries of the field 
theory.

Eq.~(\ref{fcodd}) can  also be generalized for the case $M_{L,R} \ne 0$ which
correspond to having  $6M_L+8$ semi-infinite 4-branes to the left and $6M_R+8$
semi-infinite branes to the right.  The corresponding curve is given by 
\beq
\label{fcmodd}
y~t^2~\prod_{i=1}^{M_L} (y^2-a_i^2) + e~t \sum_{j=0}^{n}
{\tilde u}_j~(y^2)^{n-j}
+f~y~\prod_{k=1}^{M_R} (y^2-b_k^2) =0.
\eeq

\mysubsection{The $SP$ case.}

Finally, we consider the case of an $SP(2M) \times U(2M+4)$ theory, 
eq.~(\ref{spgauged}). We begin with the minimal case $N_{L,R}=0$. 
 The corresponding $N=2$  brane configuration has  $6M+8$ 4-branes stretched
between  the two NS branes and four semi-infinite branes
stretching to the left ad
right  respectively. The  $N=2$ curve can now be written:
\beq
\label{csp}
v^4~t^2~+~t ~v^2~\left(~(v^2)^{3M+4} ~+~
 u_2 ~(v^2)^{3M+3} ~+~ \cdots ~ +~u_{6M+8}
\right) ~
+~v^4~=~0~.
\eeq
Imposing the $\IZ_3$  symmetry  and setting appropriate terms to zero gives
\beq
\label{cspint}
v^4~\left[t^2~+~t~ \left(~(v^2)^{3M+3}~+~ {\tilde u}_1 ~(v^2)^{3M} ~+~ 
\cdots  ~+ ~\tilde{u}_{M+1}~\right)~ +~1~ \right]=~0~.
\eeq
Notice that  the curve has factorized  with two components.  Generically  in 
moduli space  the  $SP(2M) \times  U(2M+4) $ symmetry is broken to 
$SU(4) \times U(1)^M$. In
analogy with  the $SO$ even case we expect the second component
to be relevant in 
describing the  spectral curve for the $U(1)^M$  group. 

Discarding an overall factor of $v^4$ and substituting $y \equiv v^3$ then
gives:
\beq
\label{cspf}
t^2~+~t~\left(~(y^2)^{M+1}~+~ {\tilde u}_1 ~(y^2)^{M}  ~+~ \cdots ~+~
{\tilde u}_M~(y^2)  ~ +~
{\tilde c}~ \right)~  +~ 1~=~0 ~.
\eeq
Notice that in going from eq.~(\ref{cspint}) to eq.~(\ref{cspf})
we have replaced 
the parameter, ${\tilde u}_{M+1}$, by a constant, ${\tilde c}$.  
Without this 
parameter the curve has $M$  moduli  as required from the field theory
analysis.   Semiclassically, an analysis  analogous to that in the $SO(2k+1)$ 
case above,  establishes that  ${\tilde u}_{M+1}=0 $. 
This allows for  ${\tilde u}_{M+1}$ to be  a  constant 
(dependent  on the two strong coupling scales) in the full 
quantum theory.  In fact  one finds that  generically the curve
(\ref{cspf}) has  $M+1$ holomorphic differentials,
odd under $y \rightarrow -y$;
this  would imply in turn $M+1$  photons.  
The field theory  and Type $IIA$ analysis showed that we  
should expect only $M$ photons.  The discrepancy is 
resolved by  setting  ${\tilde c}=-2$; two branch points now
coincide reducing the 
number of allowed photons to $M$ as required. 
The curve for the $Sp(2M) \times U(2M+4)$ theory  is thus  finally  given by: 
\beq
\label{fcspc}
t^2~+~t~\left(~(y^2)^{M+1}~+~ {\tilde u}_1 ~(y^2)^{M} ~ +~ \cdots ~+~
{\tilde u}_M~(y^2) 
~ - 
~2 \right)~  +~ 1~=~0~. 
\eeq
The   reader might have noticed that in above discussion, fixing the parameter
 ${\tilde c}$   is very  similar  to  that involved  in  determining the
$N=2$ $SP$ curves \cite{LLL}. 
 It  would be nice to understand these  parallels  better.  

The dependence of the curve (\ref{fcspc})  on the strong coupling scales can be
made manifest. After  rescaling $t$ and $y$
one finds\footnote{With $N_{L,R}=0$,
eq.~(\ref{spgauged}), we have  $b_0^{SU} = 4 M + 5$, $b_0^{SP} = 2 M+1$.  }
we find: 
\beq
\label{fcspscale}
t^2~+~t~\left(~(y^2)^{M+1}~+~ {\tilde u}_1 ~(y^2)^{M}  ~+ ~\cdots 
~+~{\tilde u}_M~(y^2)   
~- ~2~ \Lambda_{SP}^{ b_0^{SP}}~ \Lambda_{SU}^{b_0^{SU} }~ \right) ~+
 ~\Lambda_{SP}^{ 2b_0^{SP}}
 ~ \Lambda_{SU}^{2b_0^{SU}}~=~0~. 
\eeq
Eq.~(\ref{fcspscale}) is consistent with all the symmetries of the $SP(2M) 
\times U(2M+4)$ theory. 

For the more general case where $N_{L,R} \ne 0$, eq. (\ref{spgauged}) the 
curves can be written in a  straightforward manner. As mentioned before,
$N_L$ and $N_R$ must both be even or odd. For the even case we get the 
curve:
 \beq
\label{spm}
t^2 \prod_{i=1}^{N_L \over 2} (y^2 -a_i^2) +~ e~t~\bigl(y^2( \sum_{j=1}^{M} 
{\tilde u}_j~(y^2)^{M-j} ) -{\tilde c} \bigr) + ~f~ \prod_{k=1}^{N_R \over 2}~ 
(y^2 -b_k^2)~=~0.
\eeq
The constant ${\tilde c}$ now depends on the  values of $a_i,b_j$ as well. 
For the  odd case the curve is:
\beq
\label{spmtwo}
t^2 ~y~\prod_{i=1}^{L} (y^2 -a_i^2) + ~ e ~ t~
y^2( \sum_{j=1}^{M}    
{\tilde u}_j~(y^2)^{M-j} ) ~ + ~f~y~ \prod_{k=1}^{R }~ 
(y^2 -b_k^2)~=~0.
\eeq

We end this section with one comment  about  the
 anomalous $U(1)$ symmetries. The reader might recall 
 that the $U(1)$ subgroup of the $U(N)$ gauge symmetry in the 
$SO$ case eq.~(\ref{sogauged}), and the $U(1)$ subgroup of 
the $U(2M+4)$ gauge  
symmetry in the $SP$ case, eq.~(\ref{spgauged}), are  anomalous. 
As was discussed above, this anomaly 
is cancelled by the shift of an axion field $a$. The axion supermultiplet,
$\phi= \zeta + ia$, thus
couples to the gauge kinetic term with a coupling $\phi W^2$. In effect,
this coupling makes the strong coupling scales of the two groups,
$\Lambda_{SU}$ and $\Lambda_{SO,SP}$,  $\phi$ dependent. However,
 in the $SO$ 
case, we see that only a  
combination, $\Lambda_{SU}^{2 b_{0}^{SU}} \Lambda_{SO}^{b_0^{SO}}$ 
invariant under the anomalous $U(1)$  appears in the curve
eq.~(\ref{fcseven}), (\ref{fcsodd}). 
Thus, the $\phi$ dependence drops out of the spectral curves,
eq.~(\ref{fcseven}), (\ref{fcsodd}).   A similar argument 
shows that the curve is $\phi$ independent in the $SP$ case as well.   

The spectral curves discussed here are independent of the FI term for the 
anomalous $U(1)$. This follows from the above argument. 
The FI term gets a contribution from the 
vev of $\zeta$. Since the spectral curve is  holomorphic, $\zeta$ can only 
appear in it through the combination $\phi$. The $\phi$ independence then 
means that the curve is completely independent of $\zeta$ and thus of 
the FI term.

It was argued in ref.~\cite{DM}
 that  giving a vev to $\zeta$ corresponds to 
blowing up the orbifold and  replacing it by an ALE space. 
The argument above for $\zeta$ independence then also  implies that in  M 
theory the curves should 
stay the same when the orbifold is replaced by an ALE space. 
In the next section we see that this is indeed the case. 

\mysection{The Hyperelliptic Curves in ALE Space Backgrounds.}

In this section, we consider the effects of replacing the orbifold background
with an ALE space and show that (in appropriate coordinates) it leaves the
spectral curves unchanged.

We first discuss an orbifold background without an orientifold 4-plane.
Towards
the end we will comment on the orientifold case as well.   Consider the case
of a brane configuration, Fig.~1,   placed in a $\IC^2/\IZ_M$ background,
without any semi-infinite 4-branes.  This configuration was studied
in \cite{lpt};  for
completeness we summarize the main results here. The field theory corresponding
to placing the brane configuration in a $\IC^2/\IZ_M$ background was shown to
be an $SU(N)^M$ gauge theory  with matter
consisting of bifundamentals of two adjacent $SU(N)$ groups.  This   theory is 
in the Coulomb phase and the spectral curve
can be written down from M theory considerations. 
To  establish notation we note that the orbifold group
is a spacetime symmetry  which acts on  $v=X^4+iX^5$
and $w=X^8+iX^9$  as:
\beq
\label{defzm}
(v,~w)~ \rightarrow~ (\alpha~v,~ \alpha^{-1}~w), ~ 
\alpha \equiv e^{i {2 \pi    \over M}}.
\eeq
The orbifold can be represented as a hypersurface in  $\IC^3$. After choosing
coordinates, $y=v^M$, $z=w^M$,  and $x=vw$, it is  given by:
\beq
\label{defhyp}
y~z~=~x^M.
\eeq
Note that this surface has a singularity at  $y=z=x=0$.

The spectral  curve  for  the $SU(N)^M$ theory 
(with $t={\rm exp}(-({X^6+iX^{10} \over R}))$ is now found to be:
 \beq
\label{orbc}
t^2~+~ t~\left( y^N~+~u_1~y^{N-1}~+~ \cdots~+~u_N \right)~ +~1~=~0~.
\eeq
Reinstating the $\Lambda$ dependence  in eq.(\ref{orbc}) gives, after suitable 
rescalings,  the curve:
\beq
\label{orbcscale}
t^2~+~t~\left( y^N~+~u_1~y^{N-1}~+~ \cdots~+~u_N \right)~+~
\prod_{a=1}^M \Lambda_a^{b_0} ~=~0.
\eeq

 One can now show, by an argument analogous to that in the previous section  
that this curve is independent  of the FI terms for the anomalous  $U(1)$s. 
   In this case, 
there are $M-1$ anomalous $U(1)$s,
their FI terms correspond to the $M-1$ blow-up modes of the orbifold. However,
it turns out that the  product of  $\Lambda^{b_0}_a$  appearing  in 
eq.(\ref{orbcscale}) is  invariant under the anomalous $U(1)$s . 
Thus,   on replacing the orbifold by  a non-singular ALE space
 we expect that the curve, eq.(\ref{orbc}) should stay the same.
We turn to verifing this
now.  

  Repairing the orbifold 
singularity  corresponds to replacing it  by an ALE space, more
specifically for the
$\IZ_M$ case under discussion here, a multi-center 
Eguchi-Hanson  gravitational
instanton, described by the metric \cite{EGH}:
\begin{eqnarray}
\label{defale}
d s^2~&=&~V^{-1}~(~d \tau~+~{\vec A} \cdot {\vec dx})^2~+
~V~d x^2  \nonumber \\
V~&=&~\sum_{i=1}^{M}~{1 \over \arrowvert {\vec x} -{\vec x_i} \arrowvert}
                       \nonumber \\
{\vec  \nabla} V~&=&-    {\vec \nabla} \times  A, 
\end{eqnarray}
with $\tau$ being an angular variable and $x$ labeling points in $R^3$.
By taking  the limit  $x \rightarrow \infty$ it is easy to see that
(\ref{defale})
degenerates  to  the metric on $\IC^2/{\IZ_M}$ as required.

We will not use the full structure of  this metric here. With a suitable 
choice of complex structure, the ALE space can be described
as a hypersurface in $\IC^3$  governed by the equation:
\beq
\label{cdef}
y~z~=~\prod_{i=1}^{M}~( x - e_i ) ~,
\eeq
The coordinates $y,z$ are such that asymptotically, far from the orbifold
singularity (for large $y,z$),
$y \rightarrow v^M$, $z \rightarrow w^M$.
Comparison with eq.~(\ref{defhyp}) shows that in eq.~(\ref{cdef}) the orbifold
 singularity has been resolved by a complex structure deformation.

We would now like to show that the spectral curve does not change
on replacing the orbifold, eq.~(\ref{defhyp}), by the ALE space,
eq.~(\ref{cdef}).  In M theory,  the spectral curve
corresponds to the world volume of a single NS brane  whose
 worldvolume  has infinite extent in
$X^0,X^1,X^2,X^3$ and spans a two dimensional surface in 
 $t$, $y$, $z$, and $x$. 
This two dimensional surface
can be described by two   equations, in addition to (\ref{cdef}), involving
$t,y,z,x$.
Asymptotically,  $w \rightarrow 0$ along the
5-brane  world volume.  This boundary condition can be implemented simply
by identically setting
\beq
\label{eqtwo}
z~=~0
\eeq
on the brane world volume. 
For consistency with eq.~(\ref{cdef}), we also set  
\beq
\label{xset}
x~=~e_a~,
\eeq
where $e_a$ is one of the roots
of  the polynomial  in eq.~(\ref{cdef}).
This still leaves $y$ and $t$ undetermined
and we need one relation between them.
This equation is simply eq.~(\ref{orbc}) again.  
As in the orbifold case, it has the correct 
asymptotic behavior and also correctly describes  various other  features
of the brane configuration. Thus,  as promised above, 
we find that in going from the orbifold to the ALE space background, 
the M theory curve (in the coordinates $y,z$, eq.~(\ref{cdef})) stays the same.

One observation is worth making  at this stage.  The  ALE space,
eq.~(\ref{defale})
has $M-1$ non-trivial  two cycles.
 One can have an NS brane which wraps around one of these 
two cycles\footnote{Such a configuration can be conveniently
seen if the orbifold singularity is resolved by a K\" ahler deformation.
For the
ALE space, eq.~(\ref{defale}), the K\" ahler deformations are equivalent to
complex structure deformations.}.
If such a wrapped NS brane is brought close to the curve, eq.~(\ref{orbc}),
additional states can get massless.   In our description of the curves above
we did not have to confront this  fact  since  the curve,
eq.(\ref{orbc}),  did not 
contain a component consisting of such  wrapped NS branes.  However,
these can be present in general.  For example, 
  after     adding  extra six  branes  to the brane configuration described above 
the theory can  be in a  Higgs phase 
(besides the Coulomb phase 
analysed above).  The  transition from the Coulomb  branch
to the Higgs branch involves bringing the six branes down to 
the orbifold singularity, one expects  the extra  wrapped NS branes
components to
be important in  describing this.  We leave this    and related interesting
issues
for further  study. 

We end with some comments on the orientifold case.  In this paper we have 
considered a $\IZ_3$ orientifold
and we restrict our remarks to this case.  We saw  above that 
the orientifold charge 
forces some number of 4-branes to be stuck  at the orientifold plane.  A proper
description of these branes  in M theory  leads us again to the   questions
mentioned in  the last paragraph relating to extra components of the curve
wrapping non-trivial two cycles.  We will not be able to deal with them
adequately
here.
However,  as we saw in section 4, the spectral curves actually factorized.
One
component which  yielded information about  the $U(1)$ gauge bosons was
discussed in  some detail and  we  argued that this component should stay
the same
on  blowing up the orbifold.   We can verify this explicitly  here. 

In the orientifold case,  
the brane configuration can be thought  of 
  as being placed in $\IZ_3$  orbifold
background and the effects of the orientifold 4-plane  can be incorporated by
ensuring that  the configuration is symmetric under a reflection,
eq.~(\ref{orv}). 
We now replace the orbifold by an ALE space and would like to 
see how this affects the spectral curve.
Most of the discussion for the orbifold case  goes through here as well.
The reflection symmetry, eq.~(\ref{orv}), implies that the NS brane
world volume 
is invariant under,
\beq
\label{orientisymm}
( y,~z )~ \rightarrow - (~y, z ~)~,~ x~ \rightarrow ~x~.
\eeq
Eqs.~(\ref{xset}), (\ref{eqtwo}) which determine $x$ and $z$,
are consistent with this requirement. Furthermore, by construction, the curves
(\ref{fceven}), (\ref{fcodd}), and (\ref{fcspc}) correctly incorporate the 
constraints coming from the orientifold 4-plane. 
Thus one finds that in terms of the $y,z$ variables, (\ref{cdef}),
the curves (\ref{fceven}), (\ref{fcodd}), and (\ref{fcspc})  
stay unchanged.

\mysectionstar{Acknowledgments.}
 We would like to thank  J. Blum,  M. Douglas,  J. Harvey,  K. Intriligator, 
D. Kutasov,  W. Skiba and E. Witten  for discussions.   
The research of JL and ST is supported by the Fermi National Accelerator
Laboratory, which is operated by the Universities Research Association, Inc.,
under contract no. DE-AC02-76CHO3000. 
E.P. was supported by DOE contract no. DOE-FG03-97ER40506.

\nc{\ib}[3]{ {\em ibid. }{\bf #1} (19#2) #3}
\nc{\np}[3]{ {\em Nucl.\ Phys. }{\bf #1} (19#2) #3}
\nc{\pl}[3]{ {\em Phys.\ Lett. }{\bf #1} (19#2) #3}
\nc{\pr}[3]{ {\em Phys.\ Rev. }{\bf #1} (19#2) #3}
\nc{\prep}[3]{ {\em Phys.\ Rep. }{\bf #1} (19#2) #3}
\nc{\prl}[3]{ {\em Phys.\ Rev.\ Lett. }{\bf #1} (19#2) #3}

\end{document}